\documentclass[fleqn,twoside]{article}
\usepackage{espcrc2}
\usepackage{graphicx}
\usepackage[figuresright]{rotating}
\def\nabstar#1{\nabla\kern-0.5pt\smash{\raise 4.5pt\hbox{$\ast$}}
               \kern-4.5pt_{#1}}

\def\drvstar#1{\partial\kern-0.5pt\smash{\raise 4.5pt\hbox{$\ast$}}
               \kern-5.0pt_{#1}}

\def\newline{\relax\ifhmode\null\hfil\break\else\nonhmodeerr@\newline\fi}
\def\frac#1#2{{#1\over#2}}
\def\text#1{{\hbox{\rm #1}}}

\newcommand{\beq}{\begin{equation}}
\newcommand{\eeq}{\end{equation}}
\newcommand{\bea}{\begin{eqnarray}}
\newcommand{\eea}{\end{eqnarray}}
\def\Id{ \mbox{1\hspace{-1.2mm}I} }
\def\EQ{\hspace{-2mm} &=& \hspace{-2mm}}
\def\BE{\begin{equation}}
\def\EE{\end{equation}}
\def\BA{\begin{eqnarray}}
\def\EA{\end{eqnarray}}
\def\BAN{\begin{eqnarray*}}
\def\EAN{\end{eqnarray*}}

\def\nn{\nonumber\\}

\def\tr{\mathrm{tr}}

\def\gm5{\gamma_5}

\def\det{\mathrm{det}}

\def\BE{\begin{equation}}
\def\EE{\end{equation}}
\def\BA{\begin{eqnarray}}
\def\EA{\end{eqnarray}}
\def\BAN{\begin{eqnarray*}}
\def\EAN{\end{eqnarray*}}

\def\nn{\nonumber\\}

\def\tr{\mbox{tr}}

\def\det{\mbox{det}}

\def\text#1{{\rm #1}}
\newcommand{\AmS}{{\protect\the\textfont2
  A\kern-.1667em\lower.5ex\hbox{M}\kern-.125emS}}
\title{Recent Developments of Domain-wall/Overlap Fermions for Lattice QCD
}

\author{Ting-Wai Chiu\address[PNTU]{Physics Department, National Taiwan
        University, Taipei, Taiwan 106, Taiwan.}\thanks{
        This work was supported in part by National Science Council,
        R.O.C. under the grant numbers NSC91-2112-M002-025 
        and NSC92-2112-M002-023.}
        }
       
\begin{document}

\begin{abstract}
I review the lattice formulations of vector-like gauge theories (e.g. QCD)
with domain-wall/overlap fermions, and discuss how to optimize the chiral
symmetry for any finite $ N_s $ (sites in the fifth dimension).
In this formulation, quark propagators in gauge background can be
computed efficiently through the effective 4D lattice Dirac operator.
\end{abstract}
\maketitle

\section{Introduction}

The basic idea of domain-wall fermions (DWF)\cite{Rubakov:bb,Callan:sa}
is to use an infinite set of coupled Dirac fermion fields
$ \{ \psi_s (x), s \in (-\infty, \infty) \} $ with masses
behaving like a step function $ m(s) = m \theta(s) $ such that Weyl 
fermion states can arise as zeromodes bound to the mass defect 
at $ s = 0 $. However, if one uses a compact set of masses, then
the boundary conditions of the mass (step) function must
lead to the occurrence of both left-handed and right-handed
chiral fermion fields, i.e., a vector-like theory.

Exact chiral symmetry on the lattice was pioneered by
Kaplan \cite{Kaplan:1992bt} with his proposal of domain-wall fermions
on the 5-dimensional lattice, in which the fifth dimension
(internal flavor space) is discretized with $ N_s $ sites (flavors) and
lattice spacing $ a_5 $.
Although the initial motivation was to provide a nonperturbative
formulation of chiral gauge theories, the idea turns out to be
natural and well-defined for vector-like gauge theories (e.g. QCD), with
quark fields constructed from the boundary modes with open boundary
conditions \cite{Shamir:1993zy,Furman:ky}.
Soon after Kaplan proposed DWF for chiral gauge theories,
Narayanan and Neuberger \cite{Narayanan:wx} observed that the
chiral determinant can be
written as the inner-product (``overlap'') of two fermionic many body
states of two bilinear Hamiltonians.
For vector gauge theories like QCD, the fermion determinant
is the product of a complex conjugate pair of chiral determinants,
thus it is gauge invariant, real, non-negative, and the corresponding
lattice Dirac operator for massless quarks
(i.e., the overlap Dirac operator \cite{Neuberger:1998fp})
can be represented by a finite matrix of fixed shape regardless of the
topology of the background gauge field, without undesired doubling or
any fine-tuning.
Mathematically, the overlap Dirac operator is
exactly equal to the effective 4D lattice Dirac operator (for internal
fermions dressed with pseudofermions) of DWF (with $m_q=0$)
in the limit $ N_s \to \infty $ followed by $ a_5 \to 0 $,
\bea
\label{eq:D}
D = m_0 \left( 1 + \gamma_5 \frac{H_w}{\sqrt{H_w^2}} \right) \ ,
\hspace{4mm} H_w = \gamma_5 D_w
\eea
where $ D_w $ is the standard Wilson Dirac operator plus a
negative parameter $ -m_0 $ ($ 0 < m_0 < 2 $).

For lattice QCD with DWF, in practice,
one can only use a finite number ($ N_s $) of
lattice Dirac fermion fields to set up the domain wall, thus the
chiral symmetry of the quark fields (in the massless limit) is broken.
Also, the discretization in the fifth dimension introduces $ a_5 $ into 
the theory.
Presumably, only in the limit $ N_s \to \infty $ and $ a_5 \to 0 $, the
correct effective 4D theory with exact chiral symmetry can be recovered.
Since $ N_s $ and $ a_5 $ are independent parameters, 
it might happen that even in the limit $ N_s \to \infty $,
the massless quark propagator is chirally symmetric but still
depends on $ a_5 $, and similarly for the fermion determinant.
Thus, {\it even if $ a_5 $ is an irrelevant parameter (which decouples
in the continuum limit), it is interesting to see whether $ a_5 $
can be eliminated completely at finite lattice spacing $ a $,
or whether it can be turned into a set of parameters
(``variable spacings") such that the chiral symmetry can be preserved
optimally.}

It turns out that for the conventional DWF
with open boundary conditions \cite{Shamir:1993zy}, its 
quark propagator and effective 4D Dirac operator do {\it not} possess the 
optimal chiral symmetry for any finite $ N_s $, and they depend 
on $ a_5 $ even at $ N_s = \infty $, through the Hermitian operator
$ H $ (\ref{eq:H}) in the transfer matrix.

In this talk, I focus on the topics how to construct DWF action 
for lattice QCD \cite{Chiu:2002ir,Chiu:2002kj,Chiu:2003ir} such that
the quark propagator and the effective 4D lattice Dirac
operator possess optimal chiral symmetry for any finite $ N_s $
and gauge background. 
Due to the time constraint, I could not review recent numerical 
results on lattice QCD with domain-wall/overlap fermions. 

\section{Lattice QCD with conventional DWF}  

First, we examine the action\footnote{Here we suppress
the lattice spacing $ a $, and the Dirac, flavor, and color indices.}
of conventional DWF with open boundary conditions
\cite{Shamir:1993zy,Furman:ky}
\bea
\label{eq:DWF}
&&
\hspace{-6mm}
{\cal A}_F = 
\bar\Psi {\cal D}_{F} \Psi = 
\hspace{-2mm}
\sum_{s,s'=1}^{N_s} \sum_{x,x'}
\bar\psi(x,s)
[ a_5 D_w(x,x') \delta_{s,s'}  \nn
&& \hspace{20mm}
   + \delta_{x,x'} D_5(s,s') ] \psi(x',s'),
\eea
where $ D_w $ is the Wilson-Dirac operator minus a parameter
$ m_0 $ ($ 0 < a_5 m_0 < 2 $, $ 0 < m_0 < 2 $), 
and
\BAN
\label{eq:D5}
D_5(s,s')
\hspace{-2mm} &=& \hspace{-2mm}
\delta_{s,s'} - P_{-} \delta_{s',s+1} - P_{+} \delta_{s',s-1}, \nn
P_{\pm}
\hspace{-2mm} &=& \hspace{-2mm}
\frac{1}{2} ( 1 \pm \gamma_5 ),
\EAN
with boundary conditions:
\bea
\label{eq:bc1}
P_{+} \psi(x,0) \EQ -r m_q P_{+} \psi(x,N_s), \\
\label{eq:bc2}
P_{-} \psi(x,N_s+1) \EQ -r m_q P_{-} \psi(x,1), \\
\label{eq:r}
r \EQ [m_0(2-a_5m_0)]^{-1}.
\eea
Here $ m_q $ is the bare quark mass, and the quark fields
coupling to physical hadrons
are constructed from the boundary modes:
\bea
\label{eq:q}
q(x) = \sqrt{r} [ P_{-} \psi(x,1) + P_{+} \psi(x, N_s) ] \\
\label{eq:qbar}
\bar{q}(x) = \sqrt{r} [ \bar\psi(x,1) P_{+} + \bar\psi(x,N_s) P_{-} ].
\eea

To regularize the fermion determinant,
Pauli-Villars (pseudofermion) fields 
$ \{\phi_{xs},\bar \phi_{xs} \} $
carrying the same (flavor, color, Dirac) indices of
$ \{ \psi, \bar \psi \} $ 
but obeying the Bose statistics
are introduced \cite{Frolov:ck,Narayanan:wx}. The action 
of pseudofermions $ {\cal A}_{PF} $ is exactly the same as 
(\ref{eq:DWF}) except replacing  
$ \{ \psi, \bar \psi \} $ by 
$ \{\phi,\bar \phi \} $, with
boundary conditions ($ r m_q = 1 $):
\BAN
P_{+} \phi(x,0)        \EQ -P_{+} \phi(x,N_s), \nn
P_{-} \phi(x, N_s + 1) \EQ -P_{-} \phi(x,1).
\EAN
Thus the total action of the system is
\BAN
{\cal A} = {\cal A}_G + {\cal A}_F +{\cal A}_{PF}
\EAN
where $ {\cal A}_G $ is the gluon action.

Then the fermion determinant, the effective 4D lattice Dirac 
operator \cite{Neuberger:1997bg}, and the quark 
propagator \cite{Kikukawa:1999sy} in background gauge field 
can be obtained by integrating over  
all heavy fermion fields and pseudofermion fields.
In general, the generating functional for $n$-point Green's function of
quark fields $ q $ and $ \bar q $ can be derived as \cite{Chiu:2003ir}
\BAN
\label{eq:ZW_dwf}
Z[J, \bar J ] = \frac{\int [dU] e^{-{\cal A}_G[U] } \ \det D(m_q) \
                       e^{\bar J (D_c + m_q)^{-1} J }          }
                     { \int [dU] e^{-{\cal A}_G[U]} \ \det D(m_q) }
\EAN
where $ \bar J(x) $ and $ J(x) $ are
the Grassman sources of $ q(x) $ and $ \bar q(x) $ respectively, and
\bea
\label{eq:Dc}
r D_c  
\hspace{-2mm}
&=& 
\hspace{-2mm}
\frac{1+\gamma_5 S(H)}{1-\gamma_5 S(H)}  \\
\label{eq:SH}
S(H)   
\hspace{-2mm}
&=& 
\hspace{-2mm}
\frac{1-T^{N_s}}{1+T^{N_s}} \equiv a_5 H R(a_5^2 H^2) \\
\label{eq:T}
    T  
\hspace{-2mm}
&=& 
\hspace{-2mm}
\frac{1-a_5 H}{1+ a_5 H} \\
\label{eq:H}
    H  
\hspace{-2mm}
&=& 
\hspace{-2mm}
\gamma_5 \frac{D_w}{2 + a_5 D_w } \\ 
\label{eq:Dm}
D(m_q) 
\hspace{-1mm}
\EQ 
\hspace{-1mm}
(D_c + m_q)(1 + r D_c)^{-1} \nn
\hspace{-1mm}
\EQ 
\hspace{-1mm}
[1 + r m_q + (1 - r m_q) \gamma_5 S(H) ]/2r   \\ 
\label{eq:Dcmi}
D(m_q)^{-1} 
\hspace{-1mm}
\EQ 
\hspace{-1mm}
(1- r m_q) (D_c + m_q)^{-1} + r  
\eea

{\it The most remarkable feature of DWF is the emergence of the  
quark propagator $ (D_c + m_q)^{-1} $  
which preserves all vital symmetries of its counterpart 
$ [\gamma_\mu (\partial_\mu + i g A_\mu) + m_q ]^{-1} $ in continuum,  
while the sea quark propagator $ D(m_q)^{-1} $ for internal quark loops 
is equal to $ (D_c + m_q )^{-1} $ times a constant factor $ ( 1 - r a m_q ) $ 
plus a constant $ r a $}, as shown in Eq. (\ref{eq:Dcmi}).     

In the limit $ N_s \to \infty $,
$ S(H) \to \frac{H}{\sqrt{H^2}} \equiv \mbox{sgn}(H) $, 
then the quark propagator $ (D_c + m_q)^{-1} $ 
in the massless limit ($ m_q \to 0 $) is chirally symmetric.
Consequently, the effective 4D lattice Dirac operator for
massless internal quark loops,   
$ D(0)=D_c(1 + r D_c)^{-1}=[1+\gamma_5 S(H)]/2r \equiv D $,
satisfies the Ginsparg-Wilson relation \cite{Ginsparg:1981bj}
\BAN
D \gamma_5 + \gamma_5 D = 2 r D \gamma_5 D.
\EAN
Further, taking $ a_5 \to 0 $, then $ S(H) \to \mbox{sgn}(H_w) $, 
and $ D $ is exactly equal to the overlap Dirac operator
(\ref{eq:D}). (But {\it $ a_5 $ must be nonzero at finite $ N_s $}).

Several remarks are as follows.

(i) The axial anomaly is recovered through the Chern-Simmons current
    \cite{Golterman:1992ub,Shamir:1993zy}.

(ii) In the limit $ N_s \to \infty $, the non-singlet flavor symmetry
     $ SU_L(N_f) \times SU_R(N_f) $ of $ N_f $
     massless quarks is exact at finite lattice spacing
     \cite{Shamir:1993zy}.

(iii) The effective 4D lattice Dirac operator $ D(m_q) $ is 
      exponentially-local for sufficiently smooth background gauge fields
      \cite{Hernandez:1998et,Neuberger:1999pz,Kikukawa:1999dk}.

(iv) Any quark observable in QCD can be obtained from $ Z[J, \bar J] $ 
     by differentiation, and it possesses the discrete symmetries 
     ($ C, T, P $) of its counterpart in continuum.

\hspace{2mm} (a) Quark propagator
\BAN
& & \langle q(x) \bar q(y) \rangle =
- \left. \frac{\delta^2 Z[J,\bar J]}{\delta \bar J(x) \delta J(y)}
  \right|_{J=\bar J=0} \nn
\EQ \frac{\int [dU] e^{-{\cal A}_G} \prod_{f} \det D(m_f) \                                         (D_c + m_q)_{x,y}^{-1}}
         {\int [dU] e^{-{\cal A}_G} \prod_{f} \det D(m_f) } \
\EAN

\hspace{2mm} (b) Current-current correlator
\BAN
\hspace{-4mm} 
& &  \langle \bar d(x) \gamma_4 P_{-} s(x)
            \bar s(0) \gamma_4 P_{-} d(0) \rangle  \\
\hspace{-4mm} 
&=& \hspace{-4mm} 
    \left. \frac{\delta}{\delta J_d(x)} \gamma_4 P_{-}
             \frac{\delta}{\delta \bar J_s(x)}
             \frac{\delta}{\delta J_s(0)} \gamma_4 P_{-}
             \frac{\delta}{\delta \bar J_d(0)}
             Z[J, \bar J ] \right|_{0}  \\
\EQ
    \frac{\int [dU] e^{-{\cal A}_G } \prod_f \det D(m_f) O_K(x)}
         {\int [dU] e^{-{\cal A}_G} \prod_f \det D(m_f)}
\EAN
where
$$
O_K(x)
= -\tr [(D_c + m_d)^{-1}_{0,x} \gamma_4 P_{-}
       (D_c + m_s)^{-1}_{x,0} \gamma_4 P_{-} ].
$$
Note that the V-A structure of the left-handed quark currents
is preserved exactly.

(v) If one uses the effective 4D lattice Dirac operator $ D(m_q) $
    (or any GW Dirac operator) to define the quark action, 
    then subtleties might emerge. 
    (Recall that CP in chiral gauge theories with GW fermion is
    broken by $ O(a) $ effect \cite{Hasenfratz:2001bz}).

Consider the massless ($ m_q = 0 $) case, 
\BAN
\hspace{-2mm}
&& 
\hspace{-2mm}
{\cal A}_F = \sum_{x,y} \bar q(x) D_{x,y} q(y) \equiv \bar q D q \\
\hspace{-2mm}
&=& 
\hspace{-4mm}
\sum_{x,y} [ \bar q(x) P_{+} D_{x,y} \hat P_{-} q(y) +
                 \bar q(x) P_{-} D_{x,y} \hat P_{+} q(y) ]
\EAN
where 
$ \hat{P}_{\pm} = \frac{1}{2}[ 1 \pm \gamma_5 (1 - 2 r D) ] $ and
$ P_{\pm} = (1 \pm \gamma_5)/2 $  
are the chiral projectors for the quark fields $ q $ and $ \bar q $ 
respectively.
Then the left-handed quark current does not manifest the V-A structure, e.g. 
$\bar d(x) \gamma_\mu \hat{P}_{-} s(x) \ne \bar d(x) P_{+} \gamma_\mu s(x)$.
Consequently, the effective weak Hamiltonian constructed from 
these left-handed quark currents would break the $ SU_L(2) $ gauge symmetry  
by $ O(a) $ effect \cite{Fujikawa:2002vj}.

Note that this problem cannot be resolved by re-defining the quark field as
$ \hat q = ( 1 - r D ) q = ( 1 + r D_c )^{-1} q $, since 
$ {\cal A}_F = \bar q D q = \bar q D_c \hat q $,  
where $ D_c $ becomes nonlocal as $ N_s \to \infty $.

Thus, {\it for any Ginsparg-Wilson fermion action,  
it is impossible to define local quark 
fields with exponentially-local kernel such that
the correlation function of quark fields in gauge background can be 
expressed as  
$$ \langle q(x) \bar q(y) \rangle = (D_c + m_q)^{-1}_{x,y} \ . $$
Even though $ (D_c + m_q)^{-1} $ can be computed
via $ D(m_q)^{-1} $, conceptually, the valence quark fields 
have to be defined in terms of the boundary modes of domain-wall fermions.} 

Nevertheless, the conventional DWF has its own deficiencies: 

(a) For any finite $ N_s $,
$ S(H) $ is {\it not} the optimal rational
approximation for $\mbox{sgn}(H) $ \cite{Chiu:2002ir}, thus
the chiral symmetry is {\it not} preserved optimally.

(b) Even for $ N_s = \infty $,
$ S(H) = H/\sqrt{H^2} $ still
depends on $a_5$ through $ H = \gamma_5 D_w ( 2 + a_5 D_w )^{-1} $.

In the following, I discuss how to solve these two problems.   
%
%

\section{Zolotarev optimal rational polynomial}

The deviation of $ S(H) $ from $ \mbox{sgn}(H) $
can be measured in terms of
\bea
\label{eq:sign_error}
\sigma(S) \EQ \max_{\forall {Y\ne0}}
\left| \frac{Y^{\dagger} \{ \mbox{sgn}(H)-S(H) \} Y}{Y^{\dagger} Y} \right|
\nn 
\hspace{-2mm}
&\le& 
\hspace{-2mm}
\max_{\{\eta\}} \left| \mbox{sgn}(\eta) - S(\eta) \right| \ ,
\eea
where $ \{\eta\} $ are eigenvalues of $ a_5 H $.
Using the simple identity
\BAN
\label{eq:sign_sqrt}
|\mbox{sgn}(x)-S(x)|=|1-\sqrt{x^2} R(x^2)|  
\EAN
where $ S(x) = x R(x^2) $, we can rewrite
(\ref{eq:sign_error}) as
\bea
\label{eq:sign_isqrt_error}
\sigma(S) \le \max_{\{\eta^2 \}}
\left| 1-\sqrt{\eta^2} R(\eta^2) \right| 
\eea
where $ \{ \eta^2 \} $ are eigenvalues of $ a_5^2 H^2 $.
Clearly {\it the problem of finding the optimal rational approximation
$ S_{opt}(x) $ of $ \mbox{sgn}(x) $ with
$ x \in [-x_{max},-x_{min}] \cup [x_{min},x_{max}] $
is equivalent to finding the optimal rational
approximation $ R_{opt}(x^2) $ of
$ (x^2)^{-1/2} $ with $ x^2 \in [x_{min}^2, x_{max}^2] $.}

According to de la Vall\'{e}e-Poussin's theorem
and Chebycheff's theorem\cite{Akhiezer:1992},
the necessary and sufficient condition for an irreducible
rational polynomial (where $m \ge n, \ p_i, q_i > 0$)
\BAN
r^{(n,m)}(x)=
\frac{ p_{n} x^{n} + p_{n-1} x^{n-1} + \cdots + p_0 }
     { q_{m} x^{m} + q_{m-1} x^{m-1} + \cdots + q_0 }
\EAN
to be the optimal rational polynomial of the inverse square root function
$ x^{-1/2} $, $ 0 < x_{min} \le x \le  x_{max} $ is that
$ \delta(x) \equiv 1 - \sqrt{x} \ r^{(n,m)}(x) $ has $ n + m + 2 $
alternate change of sign in the interval $ [x_{min}, x_{max}] $,
and attains its maxima and minima (all with equal magnitude), say,
\BAN
\delta(x) =  -\Delta, +\Delta, \cdots, (-1)^{n+m+2} \Delta
\EAN
at consecutive points ($ x_i, i=1,\cdots, n+m+2 $)
\BAN
x_{min} = x_1 < x_2 < \cdots < x_{n+m+2} = x_{max}\ .
\EAN
In other words, if $ r^{(n,m)} $ satisfies the above condition,
then its error
\BAN
\sigma(r^{(n,m)}) = \max_{x \in [x_{min},x_{max}]}
                    \left| 1 - \sqrt{x} \ r^{(n,m)}(x) \right|
\EAN
is the minimum among all irreducible rational polynomials of degree $ (n,m) $.

It has been shown \cite{Chiu:2002ir} that $R(x^2)$ (\ref{eq:SH}) of
the conventional DWF (i.e., the polar approximation)
does not satisfy above criterion, thus
is {\it not} the optimal rational approximation for $ (x^2)^{-1/2} $.

The optimal rational approximation for inverse square root function
$ (x^2)^{-1/2}, x^2 \in [1,b] $ was first obtained by
Zolotarev 
in 1877, using Jacobian elliptic functions.
Explicitly,
\bea
\label{eq:rz_nn}
R^{(n,n)}_Z(x^2) = d_0
\prod_{l=1}^{n} \frac{ 1+ x^2/c_{2l} }{ 1+ x^2/c_{2l-1} },
\eea
and
\bea
\label{eq:rz_n1n}
R^{(n-1,n)}_Z(x^2) = d'_0
\frac{ \prod_{l=1}^{n-1} ( 1+ x^2/c'_{2l} ) }
     { \prod_{l=1}^{n} ( 1+ x^2/c'_{2l-1} ) } \ ,
\eea
where the coefficients $ d_0 $, $ d'_0 $, $ c_l $ and $ c'_l $
are expressed in terms of elliptic functions \cite{Akhiezer:1992}
with arguments depending only on $ n $ and $ b $.
A detailed discussion of Zolotarev's result can be found in
Akhiezer's two books \cite{Akhiezer:1992}. (see also a recent
discussion \cite{Chiu:2002eh} in the context of lattice QCD.)

The first application of Zolotarev optimal rational polynomial
in lattice QCD was to approximate \cite{vandenEshof:2001hp} 
$ (H_w^2)^{-1/2} $ in the overlap Dirac operator (\ref{eq:D}),
which improves the (optimal) rational approximation
\cite{Edwards:1998yw} via the Remes algorithm\footnote{
The Remes algorithm is a numerical (iterative) scheme to implement  
the criterion of de la Vall\'{e}e-Poussin's theorem
and Chebycheff's theorem to obtain optimal rational approximation of
any continuous function $ f(x) $. However, in the case
$ (x^2)^{-1/2} $, Remes algorithm must give less
precise numerical coefficients than those computed directly
from Zolotarev's exact solution.}. 

For lattice QCD with DWF at finite $ N_s $, the
optimization problem is how to construct
a DWF action such that the operator $ S(H) $ (\ref{eq:SH})
is replaced with
\bea
S_{opt}(H)
= \left\{ \begin{array}{ll}
  \hspace{-2mm} h R_Z^{(n,n)}(h^2),   &  N_s = 2n+1, \\
  \hspace{-2mm} h R_Z^{(n-1,n)}(h^2), &  N_s = 2n,   \\
          \end{array} \right.
\label{eq:S_opt_RZ}
\eea
where $ h = H/\lambda_{min} $,
and $ b = \lambda_{max}^2 / \lambda_{min}^2 $
($ \lambda_{min} $ and $ \lambda_{max} $ are the minimum and
the maximum of the eigenvalues of $ |H| $). However, this  
could not be solved for the conventional DWF, due to the 
functional form of $ H = \gamma_5 D_w (2 + a_5 D_w)^{-1} $.

\section{Borici's variant of DWF}

For the effective 4D Dirac operator 
$ D(m_q) $ (\ref{eq:Dm}) in the fermion determinant, it turned out 
to be possible to modify the conventional DWF action such that
its $ S(H) $ is replaced by $ S(H_w) $. 

The prescription \cite{Borici:1999zw} is to replace
$ \delta_{x,x'} D_5(s,s') $ in (\ref{eq:DWF}) with
$$
\delta_{x,x'} \delta_{s,s'} +
(a_5 D_w - 1)_{x,x'} (P_{-} \delta_{s',s+1} + P_{+} \delta_{s',s-1}). 
$$
The boundary conditions (\ref{eq:bc1})-(\ref{eq:bc2})
and quark fields (\ref{eq:q})-(\ref{eq:qbar}) are the same 
except replacing $ r $ (\ref{eq:r}) with $ r = 1/(2m_0) $.
After introducing pseudofermion fields ($ r m_q = 1 $), the
fermion determinant can be evaluated as 
\bea
&& \det ( {\cal D}_F {\cal D}_{PF}^{-1} ) \equiv \det D \nn
\EQ
\det \{ m_q + ( m_0 - m_q/2) [ 1 + \gamma_5 S(H_w) ] \}.
\label{eq:det}
\eea
In the limit $ N_s \to \infty $, $ S(H_w) \to H_w (H_w^2)^{-1/2} $, 
then the fermion determinant (\ref{eq:det}) is independent of $ a_5 $.

However, unlike the conventional DWF, its quark propagator 
in gauge background \cite{Edwards:2000qv} is abnormal,  
\bea
\label{eq:qb}
\langle q(x) \bar q(y) \rangle \ne
(D_c + m_q)^{-1}_{x,y} \ ,  
\eea
thus its quark propagator as well as other quark observables
do {\it not} manifest the discrete symmetries of their 
counterparts in continuum, similar to the problem encountered 
in the Ginsparg-Wilson fermion.

\section{Optimal domain-wall fermion (ODWF)}

The problem (\ref{eq:qb}) can be solved \cite{Chiu:2003ir} by appending  
{\it two new boundary layers at $ s = 0 $ and $ s = N_s + 1 $,
both with the constraint $ a_5 = 0 $, and re-defining 
the quark fields with these new boundary modes.} 

Further, one can turn $ a_5 $ at each layer
into a parameter $ \omega_s $.
Then this set of parameters $ \{ \omega_s \} $
can be constructed such that $ S(H_w) $ is equal to $ S_{opt}(H_w) $,
the optimal rational approximation for $ \mbox{sgn}(H_w) $
\cite{Chiu:2002ir}.

The action of optimal domain-wall fermion \cite{Chiu:2003ir}
can be written as
\bea
\label{eq:ODWF}
&&
\hspace{-6mm}
{\cal A}_F = 
\bar\Psi {\cal D}_{odw} \Psi = 
\hspace{-2mm}
\sum_{s,s'=0}^{N_s+1} \sum_{x,x'}
\bar\psi_{xs}
\{ (\omega_s D_w + 1)_{xx'} \delta_{ss'}  \nn
&&
\hspace{-7mm}
  + (\omega_s D_w - 1)_{xx'}
    (P_{+} \delta_{s',s-1} + P_{-} \delta_{s',s+1} ) \} \psi_{x's'}
\eea
with boundary conditions
\BAN
\label{eq:bc1a}
P_{+} \psi(x,-1) \EQ -r m_q P_{+} \psi(x,N_s+1), \\
\label{eq:bc2a}
P_{-} \psi(x,N_s+2) \EQ -r m_q P_{-} \psi(x,0), \ r = \frac{1}{2m_0}, 
\EAN
where $ \omega_0 = \omega_{N_s+1} = 0 $, and other $ \{ \omega_s \} $
are fixed as follows. The quark fields are constructed as
\BAN
q(x) \EQ \sqrt{r} [ P_{-} \psi(x,0) + P_{+} \psi(x, N_s+1) ] \\
\bar{q}(x) \EQ \sqrt{r} [ \bar\psi(x,0) P_{+} + \bar\psi(x,N_s+1) P_{-} ] \ .
\EAN
Then the quark propagator in gauge background can be derived as 
\cite{Chiu:2003ir}
\bea
\label{eq:quark_prop}
\langle q(x) \bar q(y) \rangle = (D_c + m_q)^{-1}_{x,y}
\eea
where
\bea
D_c \EQ 2m_0 \frac{1 + \gamma_5 S_{o}(H_w)}{1 - \gamma_5 S_{o}(H_w)}, \\
S_{o}(H_w) \EQ \frac{1 - \prod_{s=1}^{N_s} T_s}
                    {1 + \prod_{s=1}^{N_s} T_s},
\ T_s = \frac{1 - \omega_s H_w }{1 + \omega_s H_w} 
\eea
Now requiring $ S_o(H_w) $ equal to the optimal rational approximation 
of $ \mbox{sgn}(H_w) $ amounts to solving the parameters $\{\omega_s \}$ 
from the equation
\BAN
S_o(H_w) = \left\{ \begin{array}{ll}
  \hspace{-2mm} h_w R_Z^{(n,n)}(h_w^2),   &  N_s = 2n+1, \\
  \hspace{-2mm} h_w R_Z^{(n-1,n)}(h_w^2), &  N_s = 2n.   \\
          \end{array} \right.
\EAN
This is equivalent to solving the roots
$ (u_s = \omega_s^{-2}, s=1,\cdots,N_s) $ from
the nonlinear equations
\bea
\label{eq:delta_Z}
\begin{array}{ll}
\hspace{-2mm}
 1-\sqrt{u} R_Z^{(n,n)}(u)=0,   & N_s=2n+1,  \\
\hspace{-2mm}
 1-\sqrt{u} R_Z^{(n-1,n)}(u)=0, & N_s=2n.  \\   
      \end{array} 
\eea
The exact solution of (\ref{eq:delta_Z}) has been obtained 
in \cite{Chiu:2002ir}, which gives
\bea
\omega_s = \frac{1}{\lambda_{min}} \sqrt{ 1 - \kappa'^2 \mbox{sn}^2
           \left(v_s; \kappa' \right) }, \ s=1,\cdots,N_s
\label{eq:omega}
\eea
Here $ \mbox{sn}(v_s;\kappa') $ is the Jacobian elliptic function
with modulus
$ \kappa' = \sqrt{1-\lambda_{min}^2/\lambda_{max}^2} $,
and argument $ v_s $ 
\BAN
\label{eq:vs}
v_s = (-1)^{s-1} M \mbox{sn}^{-1}
   \left( \sqrt{ \frac{1+3\lambda}{(1+\lambda)^3}}; \lambda' \right)
 + \left[\frac{s}{2} \right] \frac{2K'}{N_s}
\EAN
where
\bea
\label{eq:M}
\hspace{1mm}
M \EQ \prod_{l=1}^{[\frac{N_s}{2}]}
\frac{\mbox{sn}^2 \left(\frac{(2l-1)K'}{N_s};\kappa' \right) }
{ \mbox{sn}^2 \left(\frac{2lK'}{N_s};\kappa' \right) }, \nn
\label{eq:lambda}
\lambda \EQ
\prod_{l=1}^{N_s}
\frac{\Theta^2 \left(\frac{2lK'}{N_s};\kappa' \right)}
     {\Theta^2 \left(\frac{(2l-1)K'}{N_s};\kappa' \right)},
\ \lambda'=\sqrt{1-\lambda^2}, 
\eea
$ K' $ is the complete elliptic integral of the first kind with
modulus $ \kappa' $, and $ \Theta $ is the elliptic theta function.
From (\ref{eq:omega}), one has  
$ \lambda_{max}^{-1} \le \omega_s \le \lambda_{min}^{-1} $,
since $ \mbox{sn}^2(;) \le 1 $.
Note that {\it $ \lambda_{min} (\lambda_{max}) $ can be  
fixed to be the greatest lower bound (least upper bound) of the 
eigenvalues of $ |H_w| $ for the set of gauge configurations 
under investigation. Clearly the ODWF action (\ref{eq:ODWF}) is 
always ultralocal, no matter how one fixes  
$ \lambda_{min} $ and $ \lambda_{max} $.}   

It should be pointed out that the new boundary layers at $ s=0 $ and 
$ s=N_s+1 $ with $ \omega_0 = \omega_{N_s+1} = 0 $ were first introduced 
in \cite{Chiu:2003ir}, several months after the weights 
(\ref{eq:omega}) had been obtained in \cite{Chiu:2002ir}. 
The purpose of these two new boundary layers is to render 
the quark propagator satisfying (\ref{eq:quark_prop}) such that 
any quark observable manifests the discrete symmetries of 
its counterpart in continuum.  
Since $ T_0 = T_{N_s+1} = 1 $, the transfer matrix as well 
as $ S_{o}(H_w) $ are not affected by these two 
new boundary layers.    

After introducing the pseudofermion fields ($ m_q = 2 m_0 $),
the generating function for $n$-point
function of $ q $ and $ \bar q $ can be derived as \cite{Chiu:2003ir}, 
\BAN
Z[J,\bar J]
= \frac{\int [dU] e^{-{\cal A}_G[U]} \ \det D(m_q)  \
                    e^{\bar J (D_c + m_q)^{-1} J } }
         {\int [dU] e^{-{\cal A}_G[U]} \ \det D(m_q) }
\EAN
where
\bea
D_c \EQ 2 m_0 \frac{1+\gamma_5 S_{opt}(H_w)}
                   {1-\gamma_5 S_{opt}(H_w)}  \nn
\label{eq:Dm_opt}
D(m_q) 
\hspace{-1mm}
\EQ 
\hspace{-1mm}
m_q + ( m_0 -  m_q/2 ) [1 + \gamma_5 S_{opt}(H_w)] 
\eea

It has been shown that 
$ D(m_q) $ (for any $ m_q $ and $ N_s $) 
is exponentially-local for sufficiently smooth gauge 
backgrounds \cite{Chiu:2002kj}.
Note that for any finite $ N_s $, $ D(0) $ (\ref{eq:Dm_opt}) 
is exactly equal to the overlap Dirac operator with $ (H_w^2)^{-1/2} $
approximated by Zolotarev optimal rational polynomial,
(\ref{eq:rz_nn}) ($N_s$ = odd), or (\ref{eq:rz_n1n}) ($ N_s $ = even).

Since $ S_{opt}(H_w) $ is the optimal rational approximation
for $ \mbox{sgn}(H_w) $, its error is bounded uniformly 
over the entire range of $ [1, \lambda_{max}^2/\lambda_{max}^2] $, 
and is the minimum among all rational polynomials of the same degree,
\BAN
\sigma(S_{opt})
\le \frac{1-\lambda}{1+\lambda} \simeq A(b)e^{-c(b)N_s}, \ 
b = \lambda_{max}^2/\lambda_{min}^2  
\EAN
where $ \lambda $ is defined in (\ref{eq:lambda}), and 
\BAN
A(b) 
\hspace{-2mm}
&\simeq& 
\hspace{-2mm}
4.06(1) b^{-0.0091(1)} \mbox{ln}(b)^{0.0042(3)} \\
c(b) 
\hspace{-2mm}
&\simeq& 
\hspace{-2mm}
4.27(45) \mbox{ln}(b)^{-0.746(5)}
\EAN
which can be estimated by asymptotic expansion and
numerical evaluation of $ \lambda $. 

Thus for any set of gauge configurations, one can
determine what values of $ N_s $ and $ b $
(i.e., how many low-lying eigenmodes of $ H_w^2 $ should be projected out)
are required to attain one's desired accuracy in preserving the 
chiral symmetry. 

The quark propagator in gauge background 
\BAN
&& \hspace{-8mm} \langle q(x) \bar q(y) \rangle
= ( D_c + m_q )^{-1}_{x,y} \nn
\hspace{-8mm}
&=& 
\hspace{-4mm} 
\sum_{s,s'} (P_{-} \delta_{s,0} + P_{+} \delta_{s,N_s+1})
            {\cal D}_{odw}^{-1}(x,s;y,s') \times \nn 
& & \hspace{12mm} (P_{+}\delta_{s',0}+P_{-}\delta_{s',N_s+1})
\EAN
can be computed in two different ways: \\
(a) To solve the linear system of the 5D lattice Dirac operator 
\BAN
{\cal D}_{odw}(x,s;y,s') Y  
= (P_{+} \delta_{s,0} + P_{-} \delta_{s,N_s+1}) \cdot \Id 
\EAN
(b) To solve the system of the effective 4D lattice Dirac operator   
\BAN
\hspace{-4mm} 
& & \hspace{-4mm} D(m_q) D^{\dagger}(m_q) Y = \Id \\
\hspace{-4mm} 
&=& 
\hspace{-2mm} 
    \left\{ m_q^2 + (m_0^2-m_q^2/4) 
    \left[ 2 + (\gamma_5 \pm 1) S_{opt}(H_w) \right] \right\} Y  
\EAN
with conjugate gradient, where the matrix-vector product 
$ S_{opt}(H_w) \cdot Y $ can be obtained by invoking another 
conjugate gradient (inner CG). It turns out that the best algorithm
to compute $ S_{opt} \cdot Y $  
is Neuberger's double pass algorithm \cite{Neuberger:1998jk}.

At present, the scheme (b) seems to be much more efficient than
the scheme (a), in terms of the CPU time and the memory space.
Further, there are multi-shift CG algorithms available for the scheme (b) 
to compute quark propagators for a set of quark masses,  
but so far there is {\it no} multi-mass algorithms for the scheme (a). 
For a recent quenched QCD calculation via the scheme (b),
see \cite{Chiu:2003iw}.

At this point, it is instructive to point out the  
remarkable features of Neuberger's double pass: 

(i) the memory storage for the conjugate gradient is constant 
(5 vectors), independent of the degree $ n $ of the rational 
polynomial $ R^{(n-1,n)} $. 

(ii) the CPU time is almost independent of $ n $, thus  
the matrix-vector product $ R \cdot Y $ can be 
approximated to very high precision (at large $ n $) 
without noticably extra costs.  

(iii) there exists a threshold $ n_T $ such that the double pass
algorithm is faster than the single pass algorithm for $ n > n_T $ 
(where $ n_T \simeq 12-25 $ for most platforms). 

Note that the last two properties have just been unveiled  
recently \cite{Chiu:2003ub}. 
Undoubtedly, Neuberger's double pass algorithm will become  
the standard for computing 
$ R(H^2) \cdot Y $ in lattice QCD.
   
\section{Conclusions}

A new DWF action (\ref{eq:ODWF}) for lattice QCD has been constructed 
such that the quark propagator and the effective 4D Dirac operator 
for internal quark loops have optimal chiral symmetry for any
$ N_s $ and gauge background. The quark fields are 
constructed from the new boundary modes at $s=0$ and $s=N_s + 1 $ 
with $ \omega_0 = \omega_{N_s + 1 } = 0 $ such that  
the quark propagator satisfies (\ref{eq:quark_prop}) and  
any quark observable manifests the discrete symmetries of 
its counterpart in continuum. The quark propagator 
in gauge background can be computed efficiently
through the effective 4D lattice Dirac operator,  
and the dynamical quarks can be simulated similar to 
the Wilson fermion in 5D.

\section*{Acknowledgments}

I am grateful to the Organizers of Lattice 2003
for invitation and kind hospitality. 
I thank many colleagues for discussions, 
in particular, David Adams, David Kaplan, Yoshio Kikukawa,
Rajamani Narayanan, Herbert Neuberger and Steve Sharpe. 
I also thank my (former) students, Tung-Han Hsieh,
Chao-Hsi Huang and Tsung-Ren Huang, for collaborations on
numerical simulations of lattice QCD with a Linux PC cluster 
(hep-lat/0208039, currently of 64 nodes).

\end{document}